\documentclass{PoS}
\usepackage{color}
\usepackage{cite}
\title{Higher-order hadronic and heavy-lepton contributions to the anomalous magnetic moment}

\ShortTitle{Higher-order contributions to g-2}

        \author{Alexander Kurz\\
         Institut f{\"u}r Theoretische Teilchenphysik, Karlsruhe Institute of Technology (KIT), \\76128 Karlsruhe, Germany\\
         Deutsches Elektronen-Synchrotron, DESY, Platanenallee 6, 15738 Zeuthen, Germany\\
        E-mail: \email{alexander.kurz2@kit.edu}}
        
        \author{\speaker{Tao Liu}\\
        Institut f{\"u}r Theoretische Teilchenphysik, Karlsruhe Institute of Technology (KIT), \\76128 Karlsruhe, Germany\\
        E-mail: \email{tao.liu2@kit.edu}}

        \author{Peter Marquard\\
        Deutsches Elektronen-Synchrotron, DESY, Platanenallee 6, 15738 Zeuthen, Germany\\
        E-mail: \email{peter.marquard@desy.de}}

        \author{Matthias Steinhauser\\
        Institut f{\"u}r Theoretische Teilchenphysik, Karlsruhe Institute of Technology (KIT), \\76128 Karlsruhe, Germany\\
        E-mail: \email{matthias.steinhauser@kit.edu}}

\abstract{We report about recent results obtained for the muon anomalous magnetic moment. 
          Three-loop kernel functions have been computed to obtain the next-to-next-to-leading-order
          hadronic vacuum polarization contributions. The numerical result, $a_\mu^{\rm{had,NNLO}}=1.24\pm 0.01 \times 10^{-10}$, is
          of the same order of magnitude as the current uncertainty from the hadronic contributions. 
          For heavy-lepton corrections, analytical results are obtained at 
          four-loop order and compared with the known results.}

\FullConference{ Loops and Legs in Quantum Field Theory - LL 2014,\\
		 27 April - 2 May 2014 \\
		 Weimar, Germany }

\begin{document}

\section{Introduction}
The muon anomalous magnetic moment is one of the most precisely measured quantities in particle physics
(see Refs.~\cite{Melnikov:2006sr,Jegerlehner:2008zza,Jegerlehner:2009ry,Miller:2012opa} for reviews).
However, since many years there is a discrepancy of about $3\sigma$ between experimental value~\cite{Bennett:2006fi,Roberts:2010cj} 
and theoretical prediction~\cite{Hagiwara:2011af}
\begin{equation}
a_\mu^{\rm{exp}}= 116592089(63)\times 10^{-11},
\end{equation}
\begin{equation}
a_\mu^{\rm{th}}=116591828(49)\times 10^{-11}.
\end{equation}
In quantum field theory the contributions can be classified into QED, hadronic and electroweak type. 
For the pure QED part which provides the 
largest contribution, four- and five-loop results have been obtained in Ref.~\cite{Aoyama:2012wk} 
using purely numerical methods.
It is interesting to mention
that the four-loop result is of the same order of magnitude as the discrepancy between Eq.~(1.1) and Eq.~(1.2). 
Independent cross checks for some subclasses of diagrams were calculated in Refs.~\cite{Laporta:1993ds,Aguilar:2008qj,Lee:2013sx,Kurz:2013exa}. 
In particular, Feynman diagrams containing two or three closed electron loops have been computed in Ref.~\cite{Lee:2013sx} 
and diagrams with tau loops in Ref.~\cite{Kurz:2013exa} which will be discussed later. 
Hadronic contributions which are not of light-by-light type can be obtained from measurements 
of the total cross section $\sigma(e^+e^-\rightarrow \rm{hadrons})$. 
They are the origin of the largest uncertainty. Several groups have performed the corresponding 
leading order (LO)~\cite{Davier:2010nc,Hagiwara:2011af,Jegerlehner:2011ti,Benayoun:2012wc} 
and next-to-leading order (NLO)~\cite{Hagiwara:2011af,Krause:1996rf,Greynat:2012ww,Hagiwara:2003da} analysis. 
In Ref.~\cite{Kurz:2014wya}, for the first time, the 
three-loop kernel functions have been obtained and the next-to-next-to-leading-order (NNLO) 
contributions involving the hadronic vacuum polarizations have been considered.

In our calculation we use $\tt{QGRAF}$~\cite{Nogueira:1991ex} to generate the diagrams, and then $\tt{q2e}$~\cite{Harlander:1997zb,Seidensticker:1999bb}
to transform them into
$\tt{FORM}$~\cite{Vermaseren:2000nd} readable input. $\tt{exp}$~\cite{Harlander:1997zb,Seidensticker:1999bb} and $\tt{asy}$~\cite{Pak:2010pt,Jantzen:2012mw} 
are applied to perform an asymptotic
expansion to transform the multi-scale integrals into simpler ones with one mass scale employing certain mass hierarchies.
In the $\tt{FORM}$ code we apply a projector and decompose the scalar products to end up with integrals which are reduced to master
integrals using integration by parts methods with the help of $\tt{FIRE}$~\cite{Smirnov:2008iw,Smirnov:2013dia}.

\section{Hadronic contributions}
There are hadronic LO and NLO sample diagrams contributing to $a_\mu^{\rm{had}}$ displayed in Fig.~1.
Perturbative QCD fails to give a reliable estimate for the 
hadronic vacuum polarization function $\Pi_{\mu\nu}$. The traditional approach is to transform 
$\Pi_{\mu\nu}$ into an integral over the experimental measured function 
$R(s)$ using dispersion relations and the optical theorem: 
\begin{equation}
 \Pi_{\mu\nu}= i(q_{\mu}q_{\nu}-g_{\mu\nu}q^2)\Pi(q^2), ~~\Pi(q^2)=-\frac{q^2}{\pi}\int_{m_\pi^2}^\infty \frac{\rm{d}s}{s}\frac{\rm{Im}\Pi(s)}{q^2-s},
\end{equation}
\begin{equation}
R(s)=\frac{\sigma(e^{+}e^{-}\rightarrow hadrons)}{\sigma_{0}(e^+e^-\rightarrow\mu^+\mu^-)}=12\pi \rm{Im}\Pi(s).
\end{equation}
Substituting the above expressions into the LO diagram leads to  
\begin{equation}
a_\mu=\frac{1}{3}\left(\frac{\alpha}{\pi}\right)^2\int_{m_\pi^2}^\infty \rm{d}s \frac{R(s)K(s)}{s}
\end{equation}
The kernel function $K(s)$ is obtained from the one-loop vertex diagram with the photon propagator $-i\frac{g_{\mu\nu}}{p^2}$ replaced
by $-i\frac{g_{\mu\nu}}{p^2-s}$ which looks
like a massive photon with mass $\sqrt{s}$. 
Good results for $K(s)$ are obtained assuming $s \gg m_\mu^2$ which is also true at higher orders.
$R(s)$ is provided to us by the authors of Ref.~\cite{Hagiwara:2011af} and the narrow resonance
contributions like $J/\Psi$, $\Psi(2S)$ and $\Upsilon(nS)$
($n=1,\ldots,4$) are implemented using the narrow-width approximation~\cite{Hagiwara:2003da}.
Using these ingredients, we get LO as well as NLO hadronic results 
which are in good agreement with Refs.~\cite{Hagiwara:2011af,Hagiwara:2003da}.

\begin{figure}[t]
  \begin{center}
  \begin{tabular}{cccc}
    \includegraphics[scale=0.65]{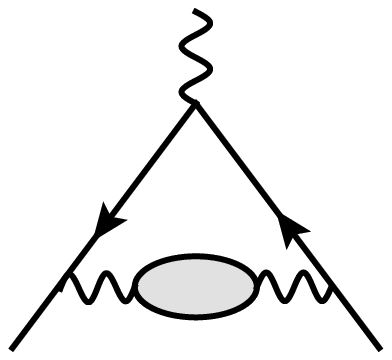} &
    \includegraphics[scale=0.65]{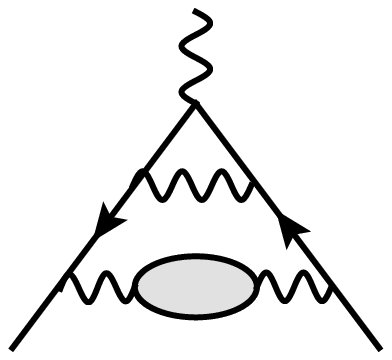} &
    \includegraphics[scale=0.65]{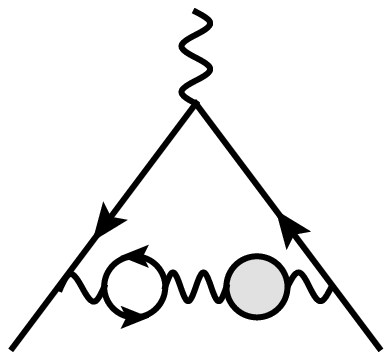} &
    \includegraphics[scale=0.65]{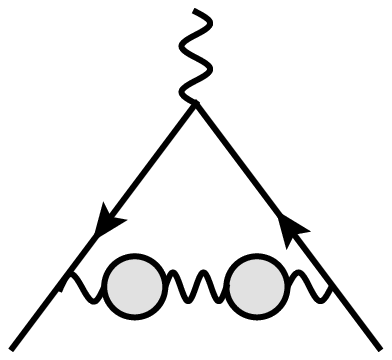}
    \\ (a) LO & (b) 2a & (c) 2b & (d) 2c \\
  \end{tabular}
  \caption{\label{fig::FD1}LO and sample NLO Feynman diagrams contributing to
    $a_\mu^{\rm had}$. The blob denotes the insertion of the hadronic vacuum polarization.}
  \end{center}
\end{figure}

At NNLO we classify the diagrams according to the number of hadronic insertions and closed electron loops. 
There are five different types denoted by $(3a),(3b),(3b,lbl),(3c)$ and $(3d)$ as shown in Fig.~2. 
Note that the contributions where the virtual electron is replaced by a tau lepton only amount to $0.01\times 10^{-10}$ at NLO, 
so we do not consider such corrections at NNLO.
Some comments are in order:
\begin{itemize}
 \item $K^{(3a)}(s)$ can be computed like the LO calculation with $s \gg m_\mu^2$ and terms up to order $(M_\mu^2/s)^4$
are obtained. Including or neglecting the highest term in $K^{(3a)}(s)$ leads to a difference at per mil level for $a_\mu^{(3a)}$  which
means a good convergence of the series. 
 \item For diagrams including electron loops, there are new asymptotic regions $m_e^2=\ell^2\ll p^2=m_\mu^2$ with 
loop momentum $\ell$ and external momentum $p$.
The non-trivial asymptotic expansion is realized with the help of the program $\tt{asy}$~\cite{Pak:2010pt,Jantzen:2012mw}, 
which is verified through
calculating the three-loop QED corrections with closed electron loops. Finally, $K^{(3b)}$ and $K^{(3b,lbl)}$ are calculated
to quartic order in $m_e$.
 \item $K^{(3c)}(s,s^{\prime})$ contains two hadronic insertions. It is calculated in various limits and 
  an interpolating function is constructed by combining the results. This
procedure was successfully tested at NLO. The final results for $a_\mu^{(2c)}$ from the exact and approximated kernels differ by 
less than $1\%$. $K^{(2c)}$ and $K^{(3c)}$ for $\sqrt{s}= 1$ GeV as a function of $\sqrt{s^\prime}$ are shown in Fig.~3.
 \item The last kernel is $K^{(3d)}$ with three hadronic insertions
\begin{eqnarray}
  K^{(3d)}(s,s^\prime,s^{\prime\prime}) &=& 
  \int_0^1 {\rm d} x \frac{x^6(1-x)}
  {\left[x^2 + (1-x) \frac{s}{M_\mu^2}\right]
    \left[x^2 + (1-x) \frac{s^\prime}{M_\mu^2}\right]
    \left[x^2 + (1-x) \frac{s^{\prime\prime}}{M_\mu^2}\right]}.
\end{eqnarray}
\end{itemize}

Analytic results for the NNLO kernels can be downloaded from the web page~\cite{progdata}. 
Inserting them into the dispersion integrals leads to the individual contributions  
\begin{eqnarray}
  a_\mu^{(3a)} &=&         0.80   \times 10^{-10} \,,\nonumber\\
  a_\mu^{(3b)} &=&        -0.41   \times 10^{-10} \,,\nonumber\\
  a_\mu^{(3b,\rm lbl)} &=& 0.91   \times 10^{-10} \,,\nonumber\\
  a_\mu^{(3c)} &=&        -0.06   \times 10^{-10} \,,\nonumber\\
  a_\mu^{(3d)} &=&         0.0005 \times 10^{-10} \,,
\end{eqnarray}
and finally to the sum
\begin{eqnarray}
  a_\mu^{\rm{had,NNLO}} = 1.24 \pm 0.01 \times 10^{-10},
\end{eqnarray}
where the uncertainty is due to the error in the experimental data.
Note that our result is of the same order of magnitude as the uncertainty of the LO hadronic contribution which is estimated to be $3.72\times 10^{-10}$ 
in Ref.~\cite{Hagiwara:2011af}. 
It is also the same order of magnitude as the experimental uncertainty anticipated for future experimental measurements~\cite{Venanzoni:2012qa}.
Thus, it should be included in the analysis of the anomalous magnetic moment of the muon. 
Note that NLO hadronic light-by-light contributions are performed in Ref.~\cite{Colangelo:2014qya}.

\begin{figure}[t]
  \begin{center}
  \begin{tabular}{cccc}
    \includegraphics[scale=0.65]{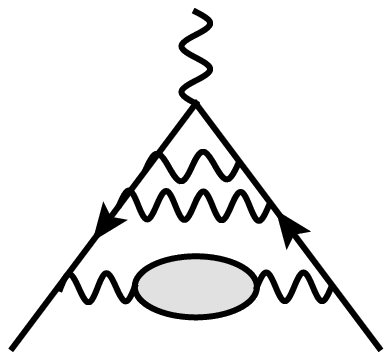} &
    \includegraphics[scale=0.65]{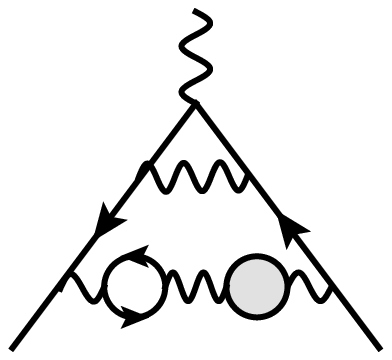} &
    \includegraphics[scale=0.65]{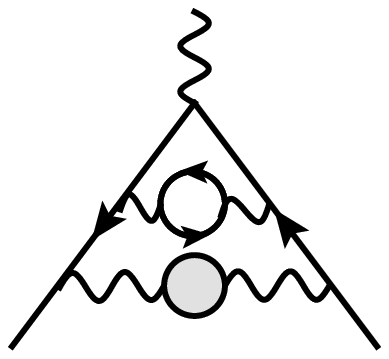} &
    \includegraphics[scale=0.65]{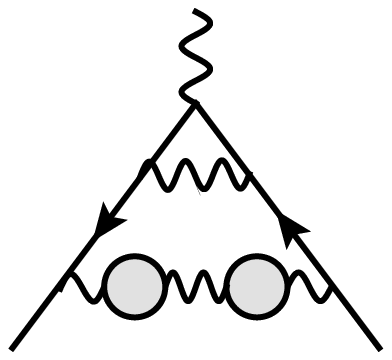}
    \\ (a) $3a$ & (b) $3b$ & (c) $3b$ & (d) $3c$ \\
    \includegraphics[scale=0.65]{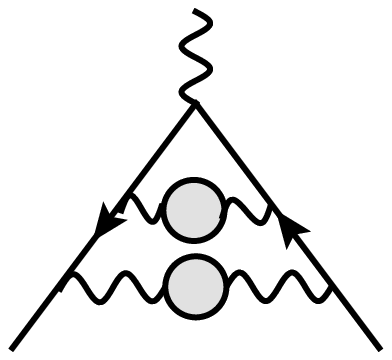} &
    \includegraphics[scale=0.65]{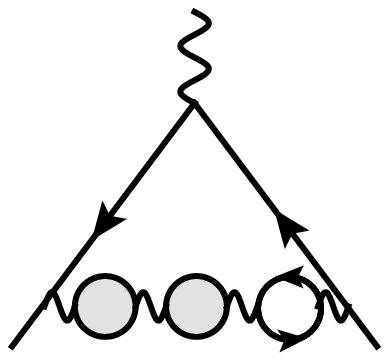} &
    \includegraphics[scale=0.65]{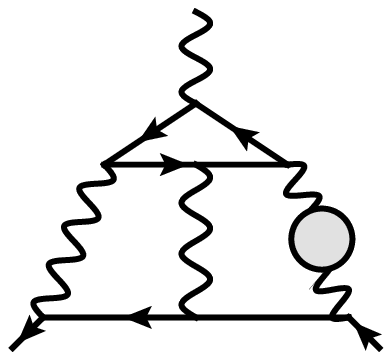} &
    \includegraphics[scale=0.65]{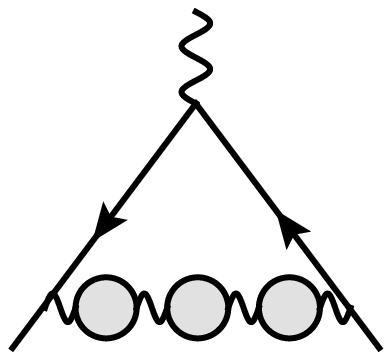}
    \\ (e) $3c$ & (f) $3c$ & (g) $3b$,lbl & (h) $3d$ \\
  \end{tabular}
  \caption{\label{fig::FD_nnlo}Sample NNLO Feynman diagrams contributing to
    $a_\mu^{\rm had}$.}
  \end{center}
\end{figure}

\begin{figure}[tb]
  \begin{center}
    \begin{tabular}{cc}
      \includegraphics[scale=0.3]{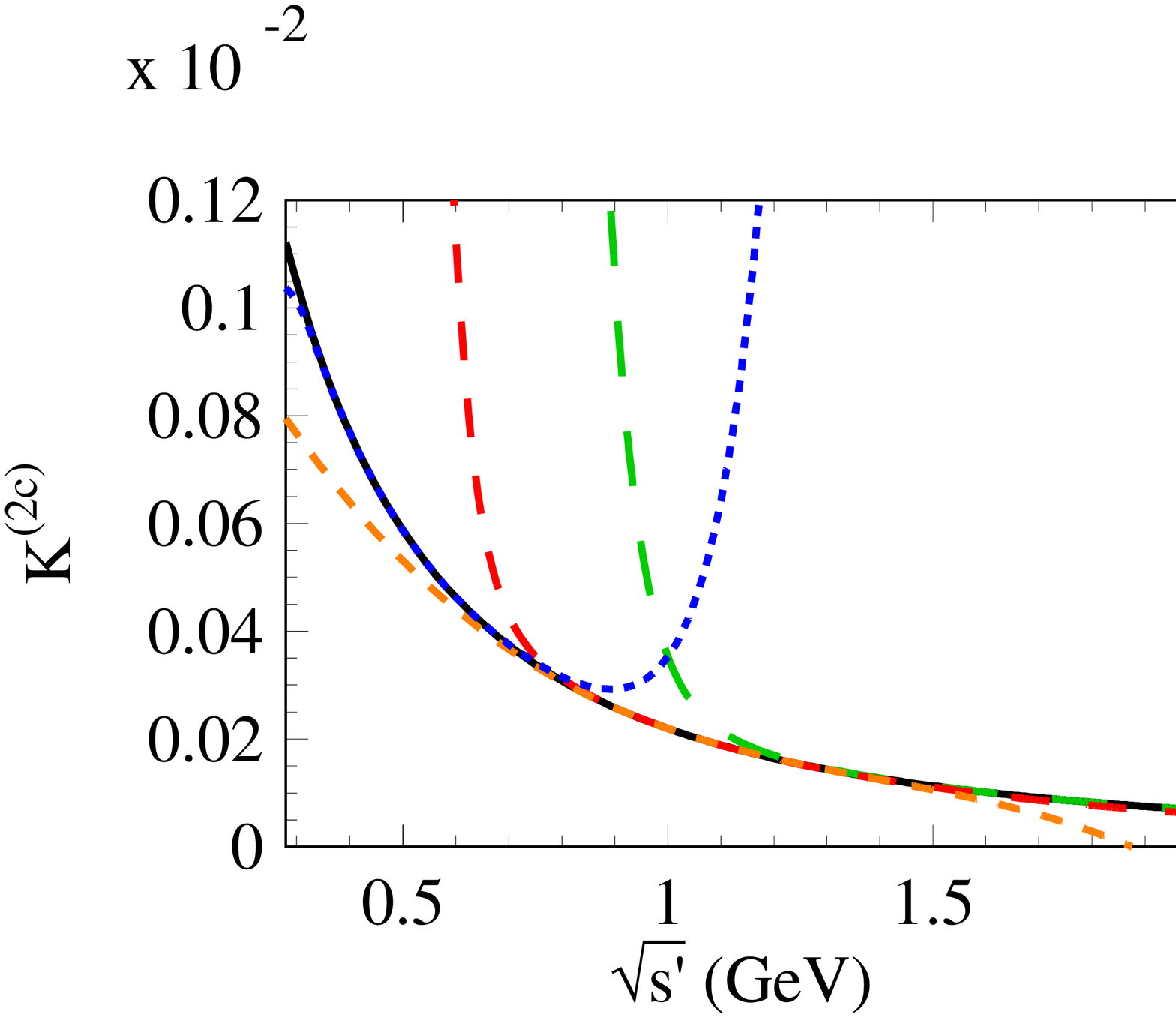} &
      \includegraphics[scale=0.3]{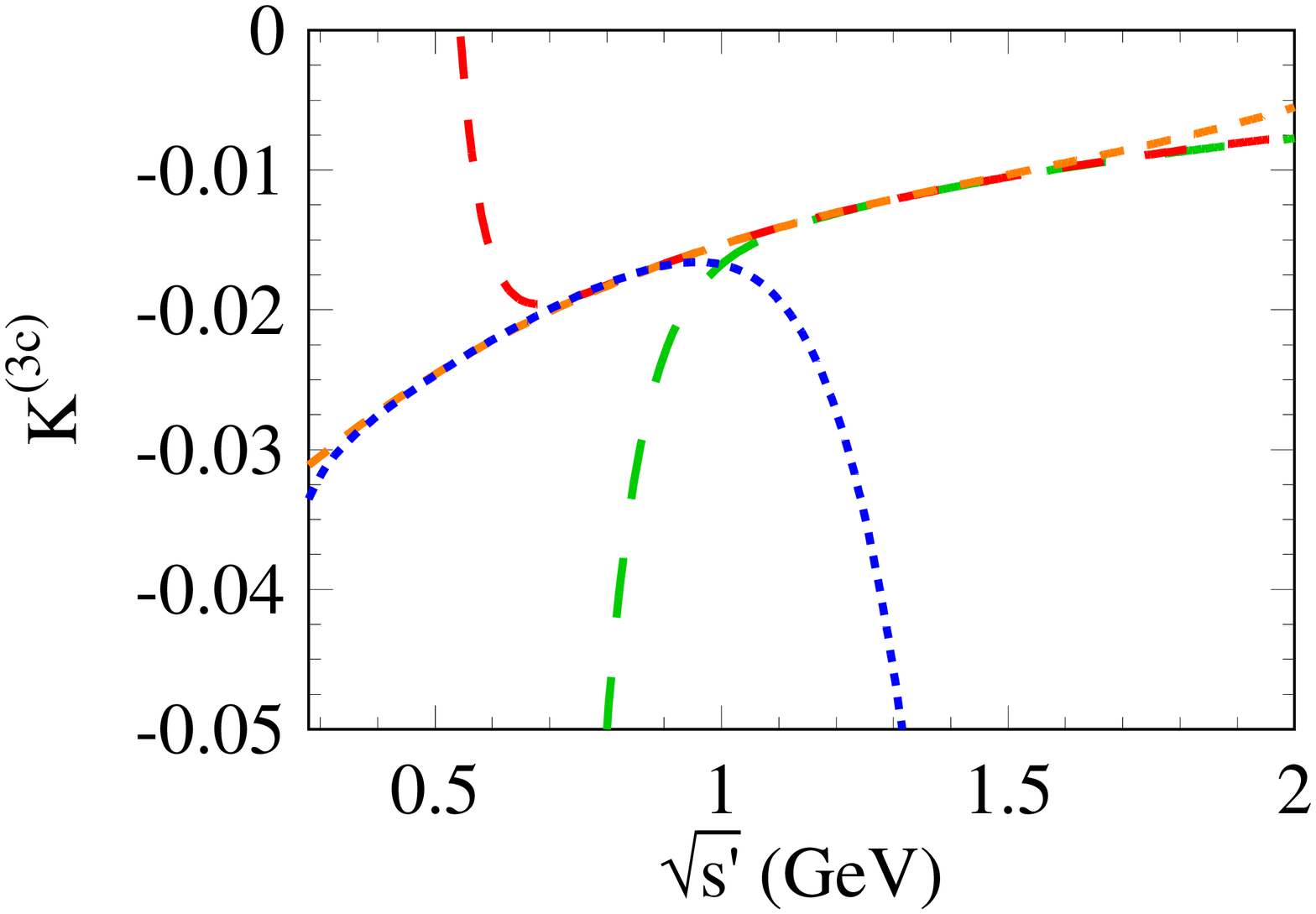}
      \\
      (a) & (b)
    \end{tabular}
    \caption{\label{fig::k2ck3c}(a) Comparison of exact result (solid, black) for
      $K^{(2c)}(s,s^\prime)$ and the various approximations for 
      $s\gg s^\prime$ (blue, dotted),
      $s\approx s^\prime$ (orange and red, short and medium dashed)
      and $s\ll s^\prime$ (green long dashed) for $\sqrt{s}=1$~GeV as a
      function of $\sqrt{s^\prime\,}$.  (b) Approximations for
      $K^{(3c)}(s,s^\prime)$.}
  \end{center}
\end{figure}

\section{Heavy-lepton contributions}
\begin{figure}
  \begin{center}
\includegraphics[scale=0.8]{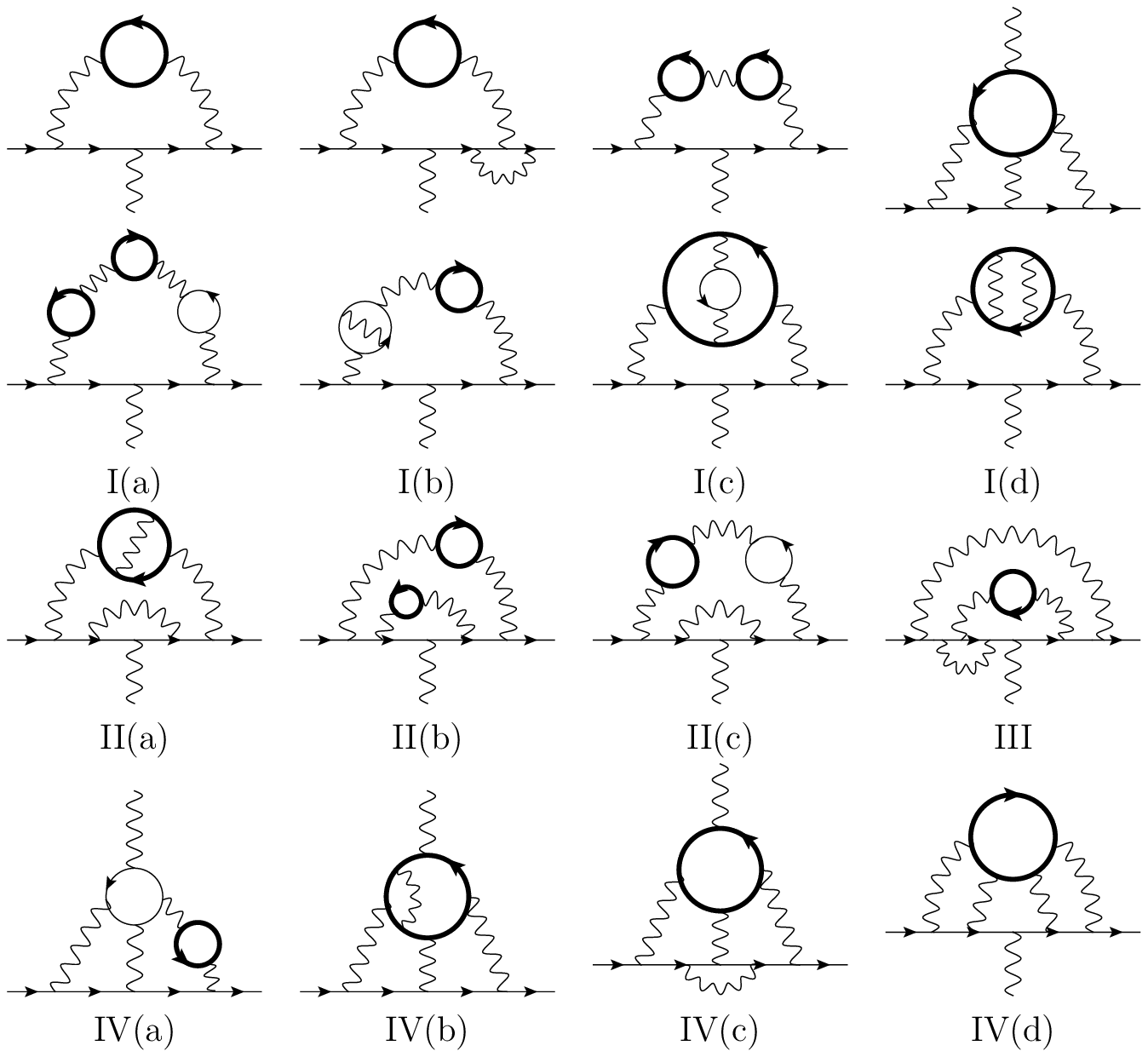}
\vspace{-11.5cm}
\end{center}
  \caption[]{Sample Feynman diagrams contributing to $a_\mu$. Thin and thick solid lines represent light and heavy leptons,
    respectively.}
\end{figure}
In Fig.~4 there are typical Feynman diagrams for heavy-lepton contribution to muon $g-2$. The symbols which label the individual diagram classes
are taken over from Ref.~\cite{Aoyama:2012wk}. These diagrams are evaluated using an asymptotic expansion with $m_\tau \gg m_\mu$. 
A graphical example for the application of the
asymptotic expansion can be found in Ref.~\cite{Kurz:2013exa}. 
During the calculation $\tt{QGRAF}$ generates 1169 diagrams at four-loop order.
Two independent calculations and a general QED gauge parameter expanded up to linear term for the leading contribution of
order $m_\mu^2/m_\tau^2$ are used to check our result.

We obtain the following analytic result 
\begin{eqnarray}
a_{\mu}(\tau) &=& 0.0078 \cdot 10^{-2} \cdot \left(\frac{\alpha}{\pi}\right)^2 + 0.0361 \cdot  10^{-2} \cdot \left(\frac{\alpha}{\pi}\right)^{3} + A_{\mu}^{(8)}(\tau) \cdot \left(\frac{\alpha}{\pi}\right)^{4}, \\
A^{(8)}_{\mu}(\tau) &=&
\left(\frac{m_\mu}{m_\tau}\right)^2 \bigg(\frac{37448693521}{2286144000}+\frac{89603}{16200}P_4
  +\frac{52}{675}P_5+\frac{4 \pi^2 \zeta_3}{15}+\frac{5771 \ln(2) \pi^4}{32400}\nonumber\\&&
  \qquad-\frac{3851 \pi^2}{3600}-\frac{25307 \zeta_5}{1440}-\frac{37600399 \pi^4}{27216000}+\frac{35590996657 \zeta_3}{508032000}\nonumber\\&&
  \qquad+\ln\frac{m_\mu^2}{m_\tau^2} \left(-\frac{38891}{12150}+\frac{19 \pi^2}{135}+\frac{3 \zeta_3}{2}\right)+\frac{359}{1080}\ln^2\frac{M_\mu^2}{M_\tau^2}\bigg)\nonumber
  \\&&
  \quad + \left(\frac{m_{\mu}}{m_{\tau}}\right)^{3} \frac{\pi^{2}}{90} + ... + {\mathcal O}\left( \left(\frac{m_\mu}{m_\tau}\right)^8 \right),
\end{eqnarray}
where $P_4=24a_4+\ln^4(2)-\ln^2(2) \pi^2$,
$P_5=120a_5-\ln^5(2)+\frac{5}{3}\ln^3(2) \pi^2$, $a_n=\mbox{Li}_n(1/2)$ and $\zeta_n$ is Riemann's zeta function.
Using $m_\mu/m_\tau=5.94649(54)\cdot 10^{-2}$ for the mass ratio, a rapid convergence of the series is observed and 
we get $A_\mu^{(8)}(\tau)\approx 4.24941(2)(53) \cdot 10^{-2}$ 
which agrees with $A_\mu^{(8)}(\tau)\approx 4.234(12) \cdot 10^{-2}$ from Ref.~\cite{Aoyama:2012wk}. In our value, the second uncertainty comes from the error in the 
mass ratio and the first one assigns to $10\%$ of the expansion terms of order $(m_\mu/m_\tau)^6$ and $(m_\mu/m_\tau)^7$. Obviously,
our result is more precise. It is interesting to note that $A_\mu^{(8)}(\tau)$ is about 100 times larger than the three-loop coefficient $A_\mu^{(6)}(\tau)$.
Inserting the numerical value for $\alpha$ we get the $\tau$-loop contribution for $a_\mu$ 
\begin{eqnarray}
10^{10} \times a_\mu(\tau)= 4.213 + 0.045 + 0.012 
\end{eqnarray}
with the numbers corresponding to two-, three- and four-loop contributions, respectively. Note that the four-loop term in Eq.~(3.3) is of the same order
of magnitude as the five-loop universal
corrections which are $0.006 \times 10^{-10}$ from Ref.~\cite{Aoyama:2012wk}. 

\begin{table}[t]
  \begin{center}
    \begin{tabular}{c||l|l}
      group & \multicolumn{2}{c}{$10^2 \cdot A^{(8)}_{\mu}(\tau)$}\\
      \hline
      & our work & old result\\
      \hline
      I(a) & \hphantom{$-$}0.00324281(2) & \hphantom{$-$}0.0032(0)\\
      I(b) + I(c) + II(b) + II(c) & $-0.6292808(6)$ & $-0.6293(1)$\\
      I(d) & \hphantom{$-$}0.0367796(4) & \hphantom{$-$}0.0368(0)\\
      III & \hphantom{$-$}4.5208986(6) & \hphantom{$-$}4.504(14)\\
      II(a) + IV(d) & $-2.316756(5)$ & $-2.3197(37)$\\
      IV(a) & \hphantom{$-$}3.851967(3) & \hphantom{$-$}3.8513(11)\\
      IV(b) & \hphantom{$-$}0.612661(5) & \hphantom{$-$}0.6106(31)\\
      IV(c) & $-1.83010(1)$ & $-1.823(11)$
    \end{tabular}
  \end{center}

\caption{\label{tab}Mass-dependent corrections to
    $A_\mu^{(8)}(\tau)$ at four-loop order as obtained in Ref.~\cite{Kurz:2013exa} and the 
    comparison with Ref.~\cite{Aoyama:2012wk}.}  
\end{table}

A detailed comparison of our result and the one from Ref.~\cite{Aoyama:2012wk} can be found in Table 1 
where uncertainties from the muon and tau lepton mass are not shown. One can see that our result based on asymptotic 
expansion provides at least two more significant digits. For group IV(b), the analytic result for the leading order expansion term
agrees with the result presented in Ref.~\cite{Kataev:2012kn}.

\section{Conclusion}
In this contribution we report about the NNLO hadronic vacuum polarization contributions to the muon anomalous magnetic moment. 
This new result reduces the discrepancy
between experimental value and theoretical prediction by about 0.2$\sigma$. We also calculated four-loop heavy leptonic corrections to $a_\mu$ in an
analytic way and find good agreement with the results in Ref.~\cite{Aoyama:2012wk}. Although not mentioned here, similar correction for $a_{e}$ are discussed in 
Refs.~\cite{Kurz:2013exa,Kurz:2014wya}.

\section*{Acknowledgements}

This work was supported by the DFG through the SFB/TR~9 ``Computational
Particle Physics''.  
P.M was supported in part by the EU Network LHCPHENOnet
PITN-GA-2010-264564 and HIGGSTOOLS PITN-GA-2012-316704.

\end{document}